\documentclass[amsmath,twocolumn,prb,aps]{revtex4-1}
\usepackage{amsthm,amsfonts,graphicx,verbatim, color, braket, dsfont, amssymb}
\usepackage{bm}
\usepackage[T1]{fontenc}
\usepackage[utf8]{inputenc}
\usepackage[colorlinks=true, allcolors=blue]{hyperref}

\newcommand{\mb}[1]{\ensuremath{\mathbf #1}}

\newcommand{\eq}[1]{Eq.~\eqref{eq:#1}}
\newcommand{\ph}{\textrm{PH}}
\newcommand{\invstiffness}{\ensuremath{\gamma}}
\renewcommand{\O}[1]{{\mathrm{O}(#1)}}
\newcommand{\U}[1]{{\mathrm{U}(#1)}}
\newcommand{\SO}[1]{{\mathrm{SO}(#1)}}
\newcommand{\Spin}[1]{{\mathrm{Spin}(#1)}}
\newcommand{\SU}[1]{{\mathrm{SU}(#1)}}
\begin{document}

\title{Half-filled Landau levels: a continuum and sign-free regularization for 3D quantum critical points}

\author{Matteo Ippoliti${}^1$, Roger S. K. Mong${}^2$, Fakher F. Assaad${}^3$ and Michael P. Zaletel${}^{1, 4}$}
\affiliation{${}^1$Department of Physics, Princeton University, Princeton, NJ 08544, USA}
\affiliation{${}^2$Department of Physics and Astronomy, University of Pittsburgh, Pittsburgh, PA 15260, USA}
\affiliation{${}^3$Institut f\"ur Theoretische Physik und Astrophysik, Universit\"at W\"urzburg, 97074 W\"urzburg, Germany}
\affiliation{${}^4$Department of Physics, University of California, Berkeley, CA 94720, USA}

\begin{abstract}
	We explore a  method for regulating 2+1D quantum critical points in which the ultra-violet cutoff is provided by the finite density of states of particles in a magnetic field, rather than by a lattice.
Such Landau level quantization allows for  numerical computations on arbitrary manifolds, like spheres, without introducing lattice defects.
In particular, when half-filling a Landau level with $N=4$ electron flavors, with appropriate interaction anisotropies in flavor space, we obtain a fully continuum regularization of the $\O5$ non-linear sigma-model with a topological term, which has been conjectured to flow to a deconfined quantum critical point.
We demonstrate that this model can be solved by both infinite density matrix renormalization group calculations and  sign-free determinantal quantum Monte Carlo.	
DMRG calculations estimate the scaling dimension of the $\O5$ vector operator to be in the range $\Delta_V \sim 0.55 - 0.7$  depending on the stiffness of the non-linear sigma model. Future Monte Carlo simulations will be required to determine whether this dependence is a finite-size effect or further evidence for a weakly first-order transition.
\end{abstract}

\maketitle

\section{Introduction}

	Understanding the space of two-plus-one dimensional conformal field theories (CFT) remains a central challenge in strongly-interacting physics.
In contrast to two-dimensions,\cite{YellowBook} comparatively little is known about the space of possible fixed points beyond large-$N$, super-symmetric, and perturbative approaches.
Where available our knowledge relies heavily on numerical Monte-Carlo simulations, and more recently, the conformal bootstrap, making it possible to compare numerical estimates of scaling exponents with rigorous analytic bounds.
A class of particular interest are the ``deconfined quantum critical points'' (DQCP) which are of  interest both to condensed matter, where they arise as Landau-forbidden phase transitions between magnetic orders with differing order parameters, and high-energy, where they are thought to provide realizations of the non-compact $\mathbb{CP}^1$ non-linear sigma model and QED$_3$.\cite{Senthil2004A, Senthil2004B, Motrunich2004, Wang2017}
While numerics support the basic picture of an emergent $\SO4$ or $\SO5$ symmetry larger than the microscopic one,\cite{Sandvik2007, Nahum2015, NahumPRX2015} 
it has proven difficult to obtain converged scaling exponents, or even conclusively determine whether the transition is a CFT\cite{Kuklov2008, Jiang2008, Nahum2015, serna2018emergence}. 
Perplexingly,  numerical estimates of the vector operator's scaling dimension appear to contradict bounds from the conformal bootstrap.\cite{Nakayama2016, Poland2018, Pufu2018}

Previous numerical studies of the DCQP considered lattice models of spins\cite{Sandvik2007, Melko2008, Lou2009, Sandvik2010, Banerjee2010, Kaul2012A, Kaul2012B, Bartosch2013, Chen2013, Shao2016, Wang2016}, 3D loop models,\cite{NahumPRX2015, Nahum2015, serna2018emergence} hard-core bosons\cite{Zhang2018} or fermions\cite{Assaad2016, Sato2017}. In these models many of the symmetries, both internal and space-time, emerge only in the IR. 
In this work we consider a  {\it continuum} regularization of the DQCP and other 3D CFTs which preserves these symmetries exactly in the UV: rather than discretizing space, the Hilbert space is made finite by Landau level (LL) quantization.
The  idea is to embed the critical fluctuations into an $N$-component ``flavor'' degree of freedom carried by itinerant fermions in the continuum. \footnote{We thank Tarun Grover for suggesting a related idea many years ago.} The motion of the fermions is  then quenched by a strong  magnetic field. When the fermions fill $N/2$ of the $N$-fold degenerate LLs (``half-filling''), fluctuations in the flavor-space give rise to a non-linear sigma model (NLSM).
This is  the famous problem of quantum Hall ferromagnetism\cite{Sondhi1993} realized experimentally both in GaAs ($N=2$) and graphene ($N=4$). In the $N=4$ case,  the resulting $\SO5$ NLSM has the  Wess-Zumino-Witten term thought to stabilize a DQCP.\cite{AbanovWiegman2000, Tanaka2005, SenthilFisher2006, LeeSachdev2015}
We  demonstrate that this  model can be studied with both  density-matrix renormalization group calculations and sign-free determinantal quantum Monte-Carlo (DQMC).

Models with exact UV-symmetries have several potential numerical advantages. 
The continuum formulation allows for the model to be defined on any manifold, such as a sphere, without introducing lattice defects. This should enable scaling dimensions to be measured using the operator-state correspondence, as well as explorations of the $F$-theorem. \cite{Klebanov2011}
Second, this realization of the DQCP has an exact $\SO4$ or $\SO5$ symmetry, whereas on a lattice it putatively emerges only in the IR at the critical point.
Because the model is essentially an explicit regularization of an $\SO5$-NLSM, it straightforward to identify the microscopic operators corresponding to the stiffness, vector, and symmetric-tensor perturbations of the NLSM. 
As such the DQCP should  exist as a \emph{phase}, e.g., without tuning, which greatly simplifies scaling collapses, and the chief question is whether the model actually flows to a CFT.
	
The paper is structured as follows.
In Sec.~\ref{sec:model} we review the model of electrons in graphene with $N=4$ flavors, its Ne\'el and valence bond solid (VBS) ordered phases, and the $\SU{N}$ anisotropies that drive the transition between them.
Sec.~\ref{sec:dmrg} contains the results of infinite density matrix renormalization group (DMRG) simulations, which are consistent with a direct, continuous transition between a N\'eel and VBS phase up to the largest system sizes.
However, our estimate of the $\SO5$ vector operator's  scaling dimension ranges from $\Delta_{V} \sim 0.55 -  0.7$ (with $2 \Delta_V = 1 + \eta$), depending on model parameters (essentially the stiffness of the NLSM). Due to the limited DMRG system size (cylinder circumference $L \lesssim 12\ell_B$), it is unclear whether this is a finite-size artifact or a signature of a weakly-first order transition.
In Sec.~\ref{sec:dqmc} we show that the model can be solved with sign-free determinantal quantum Monte Carlo, allowing for simulations with polynomial-complexity in system size, for which we present  a numerical benchmark and discuss the prospects for  large-scale simulations.
We conclude by summarizing our results and discussing future directions in Sec.~\ref{sec:discussion}.

\section{Model\label{sec:model}}
	The model is motivated by the physics of graphene in a magnetic field, where $N = 4$ flavors of two-component Dirac fermion $\Psi_a$, $a = 1, 2, 3, 4$, arise from the combination of   valley and spin degeneracy. \cite{Kharitonov2012_A, Kharitonov2012_B, Wu2014, LeeSachdev2015}
To rough approximation they are related by an $\U4$ flavor symmetry; letting Pauli matrices $\tau^\mu$ act on valley and $\sigma^\mu$ on spin ($\mu = 0$  indicates the identity), the generators are the $1 + 15$ bilinears $\tau^\mu \sigma^\nu$.
In reality the $\SU4$ part is broken down to spin $\SO3$  (generated by $\sigma^\mu$) and a near-exact $\SO2$ valley-conservation (generated by $\tau^z$).%
	\footnote{Technically, the symmetry group is $\SU2 \times \U1 \in \Spin5 \in \SU4$ which double covers $\SO3 \times \SO2$.
When restricted to an even number of particles (e.g., filling two LLs), the Hilbert space is in a bonafide representation of $\SO3 \times \SO2$.}
Microscopically, the two strongest instabilities \cite{Kharitonov2012_A, Kharitonov2012_B, Wu2014, young2014, zibrov2018} which may spontaneously break the $\SO3 \times \SO2$ symmetry are antiferromagnetism, with three-component N\'eel vector $\mathbf{N} = \tau^z \boldsymbol{\sigma}$, and the Kekule valence-bond-solid (VBS) with order parameter $e^{i \phi_{\textrm{K}}} = \tau^x + i \tau^y$ (because the valleys are at different momenta, inter-valley coherence produces a VBS distortion.)
Together these form a maximal set of anti-commuting terms $\Gamma^i = \{\tau^z \sigma^x, \tau^z \sigma^y, \tau^z \sigma^z, \tau^x, \tau^y \}$, the Clifford algebra for $\SO5$. 

For numerical purposes the Dirac fermions must be regularized, but rather than falling back to the  honeycomb lattice, we instead stick to the continuum and introduce a uniform background magnetic field $B$ orthogonal to the manifold.
The single particle spectrum collapses into $N = 4$-fold degenerate Landau levels, with energy spectrum $\epsilon_{n} = \frac{\hbar v}{\ell_B} \operatorname{sign}(n) \sqrt{2|n|}$, where $\ell_B$ is the magnetic length and $v$ is the Dirac velocity.
At zero density, the fermions should fill two of the four $n=0$ LLs, i.e., half-fill the zeroth-LL (ZLL).

When the interactions are weak compared with to the cyclotron splitting $\hbar v/\ell_B$, we can project them into the ZLL.
A phenomenological model capturing the resulting N\'eel and Kekule instabilities is an $\SU4$-symmetric contact interaction $U$ and anisotropies $u_i$,
\begin{multline}
\mathcal{H}_{\textrm{ZLL}} =  \frac{U}{2} \bigg[ \sum_{a=1}^4 \psi_a^\dagger(x) \psi_a(x) \bigg]^2 \\
  -  \sum_{i=1}^5 \frac{u_{i}}{2} \bigg[ \sum_{a, b=1}^4 \psi^\dagger_a(x) \Gamma_{ab}^i \psi_b(x) \bigg]^2.
\label{eq:Dirac}
\end{multline}
Here $\psi_a(x)$ is the field-operator of the ZLL, which can be decomposed as $\psi_a(x) = \sum_{m=1}^{N_\phi} \phi_m(x) \hat{c}_{a, m}$ for LL-orbitals $\phi_m$ on a system pierced by $N_\phi = B V / 2 \pi \ell_B^2$ flux quanta. \footnote{On  manifolds withe genus not equal to one, there is technically a ``shift'' between $N_\phi$ and the number of orbitals.}
Because each LL has one state per magnetic flux the Hilbert space is now completely finite, with $N N_\phi$ single particle states on a surface pierced by $N_\phi$ flux.

%
%
%




The anisotropies $u_i$ favor either N\'eel ($u_1 = u_2 = u_3 = u_N > 0$) or Kekule ($u_4 = u_5 = u_K > 0$) order.
A transition between the two orders is driven by the difference $u_N - u_K$, and for $u_N = u_K$ there is an exact $\O5$ symmetry (the inversion element arises from the anti-unitary particle-hole symmetry $\psi \to \psi^\dagger$).
Alternatively, taking $u_3 < u_1 = u_2$, we have an ``easy-plane'' model with at-most $\SO4$ symmetry. 

\begin{figure}
\centering
\includegraphics[width=0.99\columnwidth]{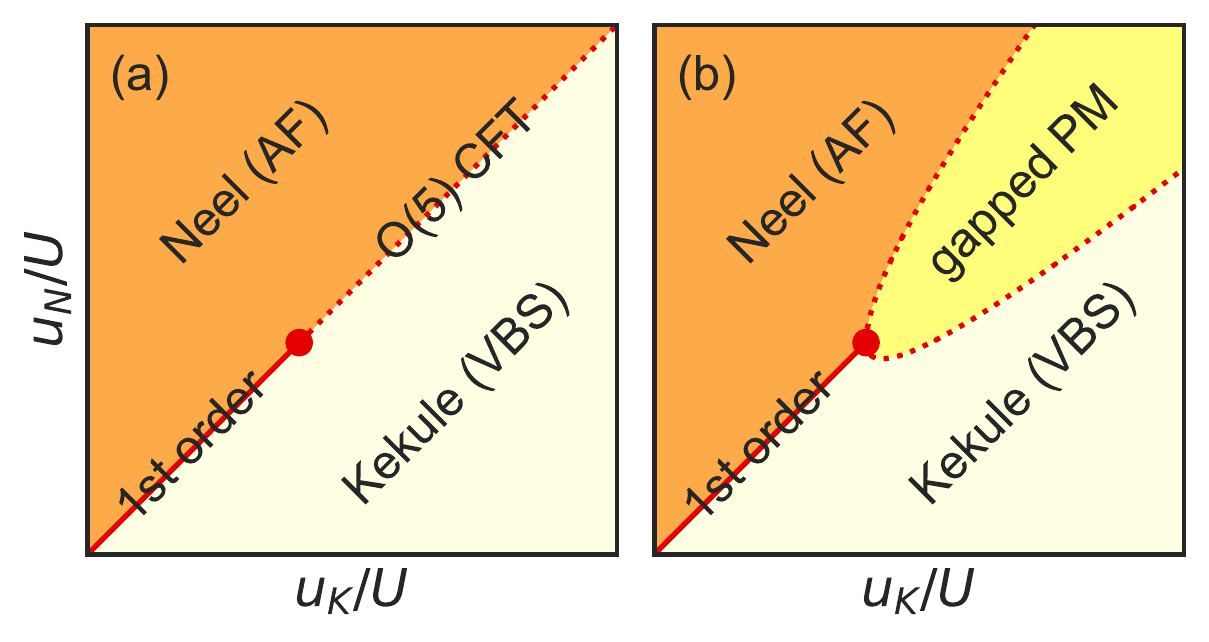}
\caption{Schematics of two possible phase diagrams of the model. 
(a) DQCP scenario. The AF and VBS orders are separated by a line ($u_N = u_K$) with manifest $\O5$ symmetry. 
For $u_i/U<w$ the symmetry is spontaneously broken, giving a first-order ``spin-flop'' transition (solid line); 
for $u_i/U>w$ there is a continuous transition characterized by an $\O5$-symmetric CFT (dotted line). This line realizes the DQCP. Alternatively, it may be that a true CFT does not exist and the transition is weakly first-order for all $U$.
(b) A Landau-allowed scenario. For $u_i/U > w$ the AF and VBS phases are separated by two independent continuous transitions (dotted lines) to a gapped paramagnetic phase, with a multicritical point at $u_N = u_K = wU$.
\label{fig:pd}}
\end{figure}

The magnetic field quenches the kinetic energy, driving quantum Hall  (anti)-ferromagnetism,  $\mathbf{n}^i = \braket{ \psi^\dagger \Gamma^i \psi} \neq 0$.\cite{Sondhi1993, Gusynin1995}
The order parameter $\mathbf{n}$ encodes which two of the four LLs are filled. 
However, in contrast to the $\SU4$ symmetric case ($u_i = 0$), where the order parameter commutes with $\mathcal{H}$ and hence doesn't fluctuate, the anisotropies lead to fluctuations.
Extending the standard $N=2$ theory of quantum Hall ferromagnetism,\cite{Sondhi1993} on the $\O5$-line these fluctuations are captured by an $\SO5$-NLSM (Euclidean) action,\cite{Wu2014, LeeSachdev2015}
\begin{align}
	S &= \frac{1}{2 \invstiffness} \int d^3 r  (\partial \mathbf{n})^2 + S_{\textrm{WZW}}[\mathbf{n}] + \dots ,
\label{eq:nlsm}
	\\	S_{\textrm{WZW}}[\mathbf{n}] &=  \frac{2 \pi i}{\operatorname{vol}(S^4)} \int\! dt\,d^3r\, \epsilon^{abcde} n^a \partial_s n^b \partial_x n^c \partial_y n^d \partial_t n^e .  \notag
\end{align}
$S_{\textrm{WZW}}$ is the $\SO5$ Wess-Zumino-Witten term, whose presence we explain shortly.\cite{LeeSachdev2015}
Note that with a magnetic field,  the particle-hole symmetry $CT$ still ensures the symmetry $\mathbf{n} \to -\mathbf{n}$.

The stiffness $1/\invstiffness$ of the NLSM is controlled by the repulsion $U$; the exchange energy from large-$U$ leads to a stiff (small $\invstiffness$) NLSM.\cite{Sondhi1993} 
Perturbing away from the $\O5$-line, $u_N \neq u_K$, will generate the ``symmetric tensor'' anisotropies $\mathcal{L}  \owns - ( \sum_i u_i \mathbf{n}^i)^2$.
Hence there is a direct correspondence between the microscopic parameters $U$, $u_i$ and the stiffness and symmetric-tensor perturbations of the NLSM respectively.

The $\SO5$-NLSM with topological term has been argued to flow to the DCQP -  unless it is too stiff, in which case the $\SO5$ may break spontaneously.\cite{Tanaka2005, SenthilFisher2006} 
So we conjecture a two-parameter phase diagram in $u_N / U$ and $u_K / U$ shown in Fig.~\ref{fig:pd}(a).
Away from the $\O5$-line, the anisotropies reduce the fluctuations and render the WZW term inoperative, so we expect N\'eel or Kekule order.
The $\O5$-line is a direct transition between the two, which may either be first or second order. If the NLSM is  stiff (small $u_i/U$)  the $\O5$ symmetry will be spontaneously broken,\cite{Wu2014} which corresponds to a first-order spin-flop transition. If the NLSM is floppy  (large $u_i/U$), the putative existence of DQCP could lead to a continuous transition which will manifest as a critical \emph{line} $u_N/U = u_K/U > u_\ast / U $ described by an $\O5$-symmetric CFT on which the scaling dimensions are constant.

	In contrast, the conventional Landau-Ginzburg-Wilson theory of phase transitions requires either a first-order transition, or two independent continuous transitions.
The two transitions will  generically be separated either by a region of phase coexistence with both N\'eel and Kekule order, or by a gapped (possibly topologically ordered) symmetric paramagnet, a possibility illustrated in Fig.~\ref{fig:pd}(b).
The transitions can only coincide when  fine-tuned to a multi-critical point.
As we will see, the numerics are in fact consistent with a direct transition along the whole $\O5$ line, though (at present)  we cannot precisely determine whether the transition at high-$u_i$ is truly continuous or just weakly first-order.

	The presence of the $S_{\textrm{WZW}}$ term can be inferred by extending the theory of $N=2$ ferromagnetic skyrmions \cite{Sondhi1993, moon1995} to $N=4$.\cite{LeeSachdev2015}
When half-filling $N=2$ flavors, it is well known that skyrmions in the ferromagnetic $\O3$ order parameter $\mathbf{n} = \psi^\dagger \boldsymbol{\sigma} \psi$ carry electrical charge.\cite{Sondhi1993}
This response is captured by the topological term  $\mathcal{L}_{\textrm{topo}} = \mathcal{A}(\mathbf{n}) \cdot \partial_t  \mathbf{n} +  \frac{\epsilon^{\mu \nu \rho}}{8 \pi} A_\mu \mathbf{n} \cdot \partial_\nu \mathbf{n} \times \partial_\rho \mathbf{n}$, where $\mathcal{A}$ is the vector-potential of a monopole and $A$ is a probe $\U1$ gauge-field.
Moving on to $N=4$, consider a skyrmion in the \emph{anti}-ferromagnetic order $\mathbf{N} = \psi^\dagger \tau^z \boldsymbol{\sigma} \psi$. 
The anti-ferromagnet has filling $\nu=1$ in each valley,  but with opposite spin. So the $N=4$ skyrmion is equivalent to an $N=2$ skyrmion in each valley independently, but with opposite handedness (due to $\tau^z$). Thus in contrast to a ferromagnetic skyrmion, the total charge is zero, but there is valley-polarization $1 - (-1) = 2 $ under $\tau^z$, the generator of the symmetry relating $\tau^{x/y}$.
More generally, we invoke $\SO5$ to conclude a skyrmion in any 3 of the 5 components induces charge under the remaining two, and a vortex (meron) under 2 of the 5 components carries spin-1/2 under the remaining three. This is the physics of $S_{\textrm{WZW}}$. 
A second consequence of anti-ferromagnetism is the cancellation of the $\mathcal{A} \cdot \partial_t \mathbf{n}$ term to leading order, with fluctuations generating $(\partial_t \mathbf{n})^2$.\cite{DasSarma1998}

\section{Infinite DMRG simulations\label{sec:dmrg}}

In this Section we study the model on an infinitely-long cylinder of circumference $L$ in order to use infinite density matrix renormalization group (iDMRG) \cite{mcculloch2008} numerical simulations.
The difficulty of the DMRG blows up exponentially with the circumference, which (relative to the UV cutoff $\ell_B$)  restricts us to  smaller system sizes ($L \sim 12 \ell_B$) than previous lattice Monte-Carlo simulations.
Nevertheless our results appear consistent with the conjectured phase diagram of Fig.~\ref{fig:pd}a).

\subsection{Method}
After projecting the Hamiltonian in \eq{Dirac} into the $n=0$ LL, the contact interactions become familiar Haldane $V_0$ pseudopotentials.
We then solve for the ground state on an infinitely-long cylinder of circumference $L$ using the iDMRG algorithm developed for multi-component quantum Hall states\cite{Zaletel2013, Zaletel2015}.
Our numerics exactly conserve the quantum numbers of charge, spin, and valley, while the rest of the $\O{N}$ symmetry becomes manifest as the numerics converge.

Infinite-cylinder DMRG has two IR cut-offs: the cylinder circumference $L$, and the finite ``bond-dimension'' $\chi$  (e.g., accuracy) of the DMRG numerics.
The latter is the dimension of the matrices used in the matrix product state (MPS) variational ansatz, which limits the amount of entanglement in the state to $S \sim \log \chi$, while the computation time goes as $\chi^3$.
By construction, an MPS with finite $\chi$ has exponentially decaying correlations, $\langle O(r) O(0) \rangle \leq  a e^{-r/\xi}$ at large $r$ for some $\xi$ called the ``MPS correlation length.''
Thus at a 1+1D critical point, where the system has algebraic correlations $\langle O(r) O(0) \rangle \sim r^{-\Delta_{O}}$ along the length of the cylinder, the MPS ansatz can only capture the power-law decay  out to a finite length $\xi(\chi)$.
This leads to the idea of ``finite-entanglement scaling'' (FES)\cite{tagliacozzo2008scaling, PollmannFES}: near a critical point, the  bond dimension $\chi$ introduces an additional $\chi$-dependent length scale $\xi$, which can then be factored into any scaling collapse.
In the present case, the putative 2+1D critical point does not actually dimensionally reduce to a 1+1D critical point on the cylinder (see below). Nevertheless, at finite $\chi$ and large $L$,  the $\xi$ of the MPS is not that of the true ground state, so we extract properties from  \emph{two}-parameter scaling collapses in $L$ and $\xi$.

 \subsection{Cylinder diagnostics of the 2D phases}
 	The 2+1D phases we wish to distinguish are: 
	(1) an ordered phase in which $\SO2$ is spontaneously broken (e.g.\ VBS, XY, or Kekule order); 
	(2) an ordered phase in which an $\SO{N}$ symmetry for $N = 3, 4$ or $5$ is  broken;
	(3) a gapped paramagnetic phase; 
	and (4) an $\SO{N}$-CFT. 
The subtlety, however, is that for fixed cylinder circumference $L$, each of 2, 3, 4 dimensionally reduces to a 1+1D gapped, symmetric paramagnet, so we must elucidate how we distinguish them within our numerics.

To do so we place the $\O{N}$-NLSM of Eq.~\eqref{eq:nlsm} on a cylinder.
If the symmetry is spontaneously broken in 2+1D,  then we can take $\partial_y \mathbf{n} \sim 0$ where $y$ runs around the cylinder, and obtain
\begin{align}
S_{\textrm{cyl}} = \frac{L}{2 \invstiffness} \int dx dt ( \partial \mathbf{n}(x, t) )^2 + \cdots
\end{align}
Here $\partial$ is the derivative in 1+1D, and the WZW term vanishes because  $\partial_y \mathbf{n} = 0$ (the skyrmions are gapped on the cylinder).
This is a 1+1D $\O{N}$  NLSM without a topological term, and with stiffness $L/\invstiffness$. 
For $N>2$ this model is gapped, with a finite correlation length $\xi_{\textrm{1D}} \sim a e^{2 \pi\frac{N}{N-2} \frac{L}{\invstiffness}}$.\cite{Sachdev_book}
For $N=2$ the system will have algebraic order, unless $L/\invstiffness$ is small enough to drive a Berezinskii-Kosterlitz-Thouless transition into a disordered phase.
Hence for cases (1) and (2), 2+1D spontaneous symmetry breaking will manifest as a $\xi_{\textrm{1D}}$ which scales exponentially with $L$ ($N>2$) or may be infinite ($N=2$).

In contrast, for case (3), a 2+1D gapped paramagnet, the $\xi_{\textrm{1D}}$ will saturate with $L$ to the true $\xi$ of the 2+1D phase.
 
Finally, for case (4) the system is a 2+1D CFT and we cannot approximate $\partial_y \mathbf{n} = 0$. Scale invariance instead dictates that $\xi_{\textrm{1D}} \propto L$, and the behavior of other observables can be determined by conformal finite-size scaling in $L$.

\subsection{Continuous transition}

\begin{figure}
\centering
\includegraphics[width=\columnwidth]{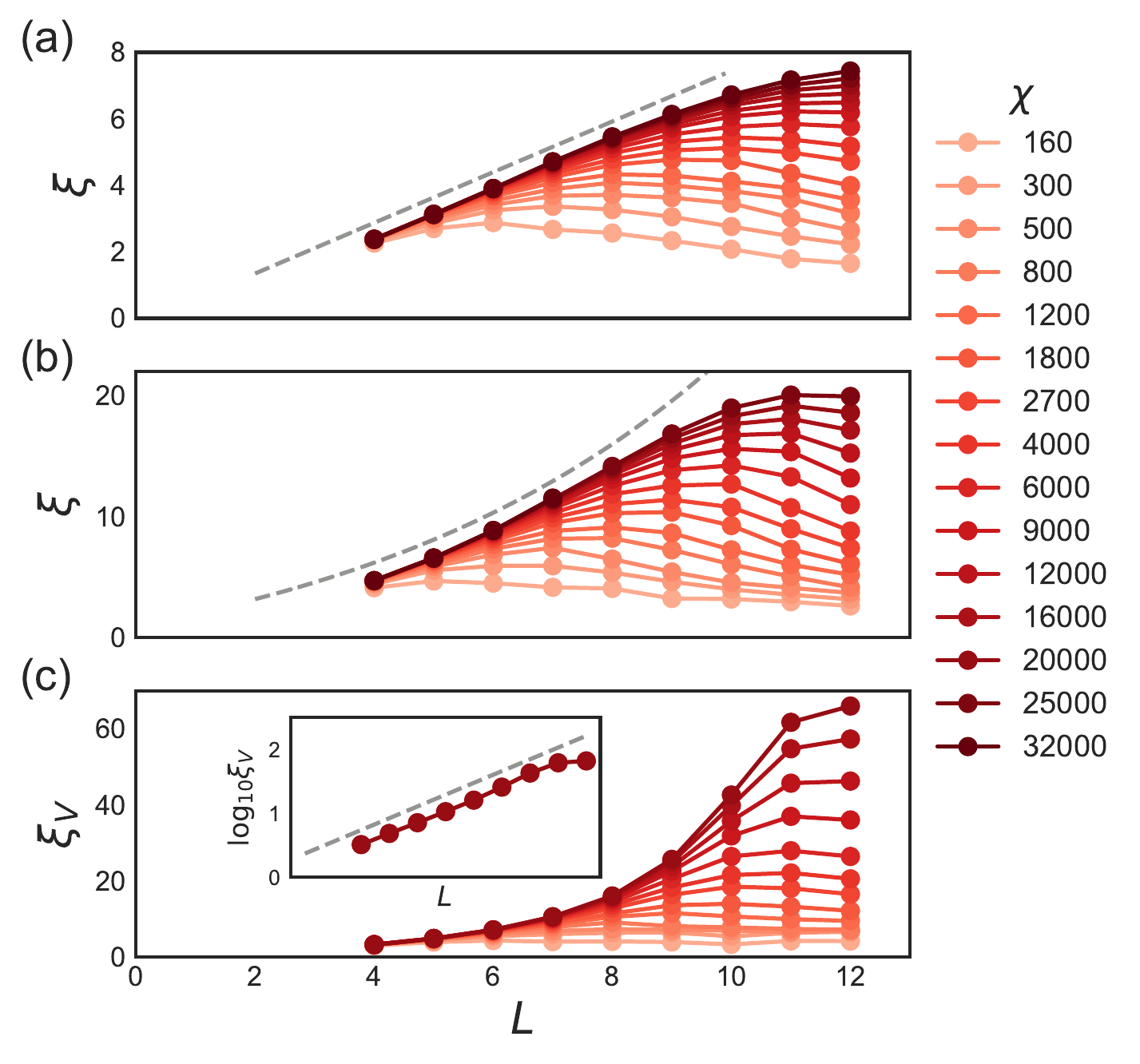}
\caption{
Correlation length as a function of cylinder circumference $L$ and bond dimension $\chi$,
obtained from numerical iDMRG simulations.
(a) On the $\O5$ line, with small stiffness: $U = 2$, $\bar{u} = u_3 = 1$, $m=0$. 
As $\chi$ is increased, $\xi$ approaches a linear dependence on size, $\xi \sim \alpha L$ (dashed line), consistent with a CFT on the cylinder.
(b) On the $\O5$ line, with large stiffness: $U = 10$, $\bar{u} = u_3 = 1$, $m=0$. 
$\xi(L)$ is concave-up (dashed line is an exponential fit to the first 4 points at largest $\chi$), consistent with a weakly-first-order transition.
(c) On the VBS side: $U=2$, $\bar{u}=1$, $u_3=0$, $m=-0.1$. 
The correlation length in the valley channel $\xi_V$ diverges exponentially with $L$ (inset shows a semilogarithmic plot), clearly indicating a symmetry-broken state.
\label{fig:xi}}
\end{figure}

	To assess the plausibility of the scenario shown in Fig.~\ref{fig:pd}(a) we first measure the scaling of $\xi_{\textrm{1D}}$ with $L$. We set $u_1 = u_2 = \bar{u} + m$ and $u_4 = u_5 = \bar{u} - m$, so that the AF-VBS transition is driven by $m$ ($m=0$ defines the critical point), while $u_3 \leq \bar{u}$ can be used to introduce easy-plane anisotropy. The repulsion $U$ sets the overall spin stiffness.
In Fig.~\ref{fig:xi} we show the correlation length $\xi$, defined by the dominant eigenvalue of the MPS transfer matrix,\cite{Schollwock2011} for several representative points. For $m=0$ and small $U$ (i.e., on the putative critical line), the scaling of $\xi \propto L$ is perfectly linear. In contrast, for  $m\neq 0$ (i.e., in an ordered phase), or for $m=0$ and large $U$ (i.e., on the putative first-order transition line), $\xi$ grows super-linearly and is well fit by an exponential dependence in both cases.
For $m \neq 0$ the the exponential form is clear over more than a decade, while  for $m = 0$, large $U$, we can really only detect a positive curvature, or concavity.

	 The linear-$L$ behavior for small $U$ is consistent with scenario Fig.~\ref{fig:pd}a), though we cannot rule out a gapped paramagnet, Fig.~\ref{fig:pd}b), with a correlation length $\xi_{\textrm{2D}} \gtrsim 12 \ell_B$ greater than the circumference we can access.
Likewise, while the super-linear behavior for large $U$ indicates a region of first-order behavior, we cannot rule out a transition which is \emph{weakly} first-order along the whole $m=0$ line.
As $U$ varies along the $m=0$ line, the curvature in $\xi(L)$ onsets smoothly, and becomes clear in our numerics for $U\gtrsim 5\bar{u}$.

\subsection{Scaling dimensions}

To investigate the intriguing possibility of a CFT in the small-$U$ regime, we attempt to measure the scaling dimension $\Delta_V$ of the vector operator $\mathbf{n}^i(r) = \psi^\dagger(r) \Gamma^i \psi(r)$.
Assuming conformal invariance, on the plane the two-point function is
$$
C_{ij}(r) = \langle  \mathbf{n}^i(r)  \, \, \mathbf{n}^j(0) \rangle \propto \delta_{ij} r^{-2 \Delta_V}
$$
where $\Delta_V$ is the scaling dimension. Since the $\SO5$ symmetry is exact in our numerics, we can restrict to a single $C_{ii}$ (for $\SO4$, $i \neq 3$).
On the cylinder, we measure $\Delta_V$ via the total squared ``magnetization'' $M_i$, \begin{equation}
M_i^2(L) \equiv \int_{\mathbb R} dx \int_0^L dy \, C_{ii}(x,y) \;.
\end{equation}
The dependence on the cylinder circumference $L$ is easily isolated via scaling collapse:
\begin{align}
M_i^2(L) =  L^{2-2\Delta_V} M_i^2(1).
\end{align}
Thus in principle  $\Delta_V$ can be extracted from $M_i^2(L)$ using a one-parameter finite-size scaling collapse.

This picture is complicated by the finite bond dimension $\chi$ in our iDMRG numerics, which as discussed earlier introduces a second length cutoff in the problem in the form of a finite correlation length $\xi$.
So we calculate $M_i^2(L)$ for a range of values of $L$ and $\chi$, with the latter parameterized via the MPS correlation length $\xi$, and collapse the data using the scaling form
\begin{equation}
M_i^2(L,\xi) = L^{2 - 2 \Delta_V} f(\xi/L) \;.
\label{eq:2param_scaling}
\end{equation}
\begin{figure}
\centering
\includegraphics[width=0.99\columnwidth]{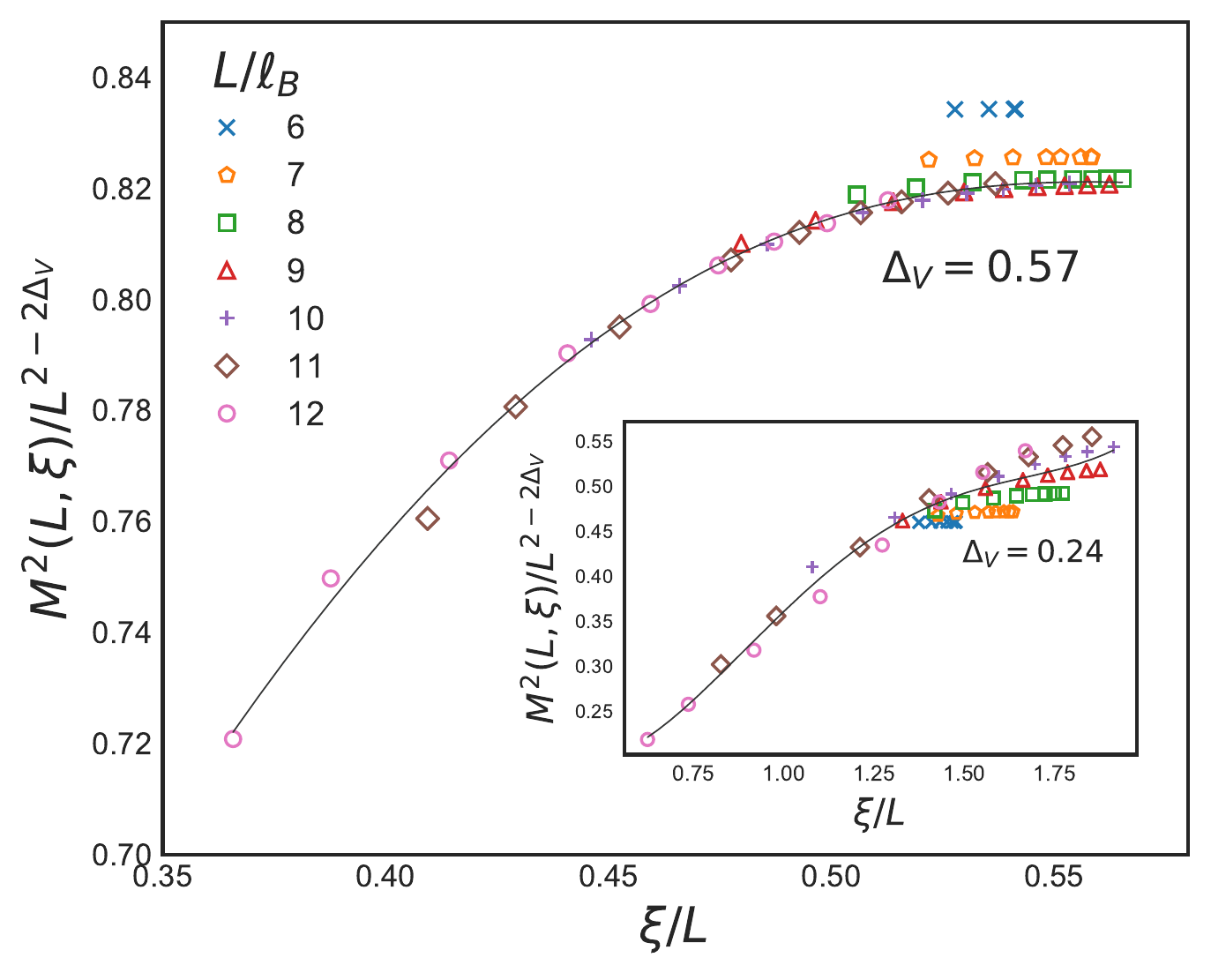}
\caption{Two-parameter scaling collapse of $M_i^2$ on the $\SO5$ line, $u_i \equiv \bar{u}$, with $U = 0.5\bar{u}$.
The data is obtained from iDMRG simulations with bond dimensions $\chi$ ranging from $2000$ to $32000$ (leftmost to rightmost points at each size).
The solid line shows a polynomial fit to data points with $L\geq 8\ell_B$ and represents the scaling function $f(\xi/L)$ in Eq.~\eqref{eq:2param_scaling}; $\Delta_V$ is chosen so as to minimize the error of the fit.
Inset: data for $U = 10\bar{u}$ (large stiffness) shows poor collapse and returns an estimate of $\Delta_V$ in severe violation of the unitarity bound, $\Delta_V > 0.5$. 
\label{fig:scaling}}
\end{figure}
For large enough circumference ($L\gtrsim 8\ell_B$), we find that there exists a value of $\Delta_V$, typically determined to within $\pm 0.01$, such that the data for different $L$ up to $12\ell_B$ collapse onto the scaling form  of \eq{2param_scaling}.
An example is shown in Fig.~\ref{fig:scaling}.
Similar behavior is found across much of the parameter space (we sit at the critical point, $m=0$, and assume $\bar{u}>0$, leaving the two independent parameters $U/\bar{u}$ and $u_3/\bar{u}$).

\begin{figure}
\centering
\includegraphics[width=\columnwidth]{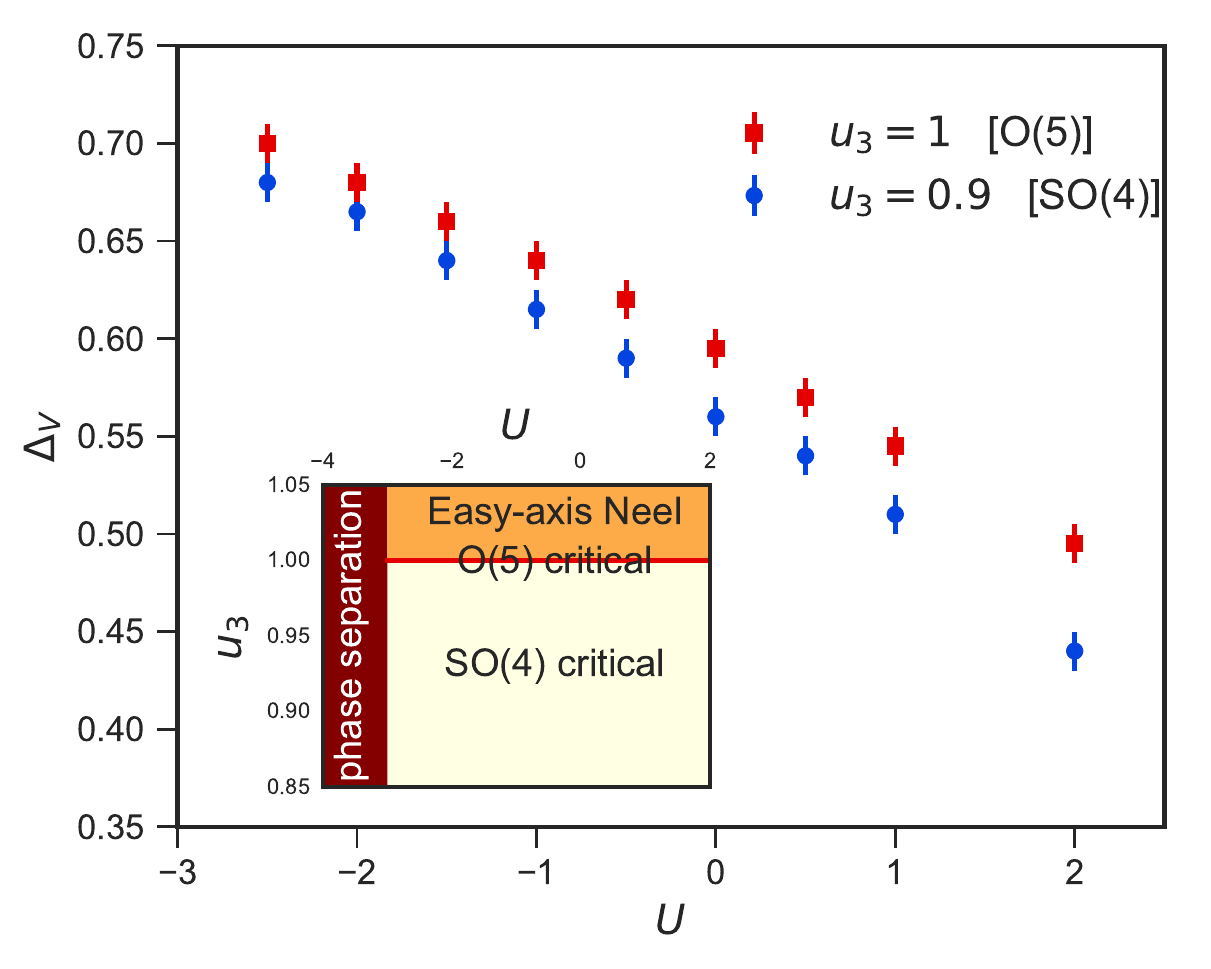}
\caption{
Measured scaling dimension $\Delta_V$ along two cuts in parameter space at $m=0$ and $\bar{u}=1$:
$u_3 = 1$ ($\O5$ symmetry) and $u_3 = 0.9$ ($\SO4$ symmetry).
Note that while we  refer to $U / u_i$ as the stiffness, making the region $U < 0$ seem unphysical, the $u_i$ themselves lead to a repulsive interaction which prevents phase-separation for $U \gtrsim -2.6$.
 For $u_3>1$, the system polarizes into an easy-axis N\'eel state. 
\label{fig:drift}}
\end{figure}

In Fig.~\ref{fig:drift} we show the variation of the estimated $\Delta_V$ along two cuts in parameter space, one on the $\O5$-line and one in the $\SO4$-region.
The value of the scaling dimension $\Delta_V$ (as well as the accuracy of the collapse) drifts with $U, u_3$, with large-$U$ having lower $\Delta_V$ and worse collapse.
The conjectured $\O{N}$-symmetric CFT should yield one well defined value for $\Delta_V$ for $N = 4, 5$ respectively, so the dependence we observe is either a finite-size artifact or evidence that a weakly-first order transition persists to higher $u_i/U$ that can be detected from the super-linear scaling of $\xi$.
Indeed, at large $U$, $\Delta_V$ violates the unitarity bound $\Delta_V \geq \frac{d}{2}-1 = \frac{1}{2}$, while lowering $U$ takes $\Delta_V$ up to $\sim 0.7$ ($U$ can be reduced down to $U \simeq -2.6\bar{u}$, at which point the attractive interaction leads to phase separation).
Due to the limited system size, it is difficult to determine where (if anywhere) the weakly first-order line becomes a CFT.

The easy-plane anisotropy ($u_3 < \bar{u}$) breaks $\O5$ down to $\SO4$ and makes the model stiffer.
A moderate value like $u_3 = 0.9\bar{u}$ (used in Fig.~\ref{fig:drift}) lowers $\Delta_V$ slightly, while a large anisotropy like $u_3 = 0$ makes the transition strongly first-order.

In conclusion, while iDMRG simulations do not provide a definitive numerical prediction for the scaling dimension $\Delta_V$, they are consistent with a continuous transition characterized by an exponent $\Delta_V$ somewhat larger than the unitarity bound, in agreement with earlier calculations on the cubic dimer model\cite{Sreejith2014, Sreejith2015}, the $JQ$ model\cite{Pujari2013, Pujari2015, Harada2013, Lou2009}, loop models \cite{NahumPRX2015} or large-$N$ expansion of the $\mathbb{CP}^{N-1}$ field theory\cite{Dyer2015}, all of which place the vector dimension $\Delta_V$ in the range $0.57$ to $0.68$.

\section{Sign-free Determinantal Quantum Monte Carlo\label{sec:dqmc}}
We now show that the model is amenable to sign-free determinantal quantum Monte Carlo,  due to a combination of particle-hole and flavor symmetry, leading to an algorithm with polynomial complexity in system size.

We consider a quantum Hall Hamiltonian of the general form
\begin{align}
H &= \frac{1}{2} \sum_i \int d^2 r \, \, n^i(\mb{r}) U^i(\mb{r} - \mb{r}') n^i(\mb{r}') \\
&=  \frac{1}{2 V} \sum_i  \int d^2 q  \,\, n^i_{-\mb{q}} U^i( \mb{q}) n^i_{\mb{q}}
\end{align}
in real and Fourier space respectively ($V$ is volume). On the sphere, the Fourier transformation can be replaced by a spherical harmonic decomposition.
Here  $n^i(\mb{r}) =  \psi^\dagger(\mb{r}) O^i \psi(\mb{r})$, where $O$ acts on the flavor index. Without loss of generality we take $O = O^\dagger$, so that $n$ is Hermitian. 
After LL-projection on a circumference $L$ cylinder in Landau gauge, the single particle  orbitals  are labelled by their momenta around the cylinder, $k = \frac{2 \pi}{L} m$ for $m \in \mathbb{Z}$, and flavor index $a$.
The density operators are expanded in annihilation operators $\hat{c}_{k, a}$ as \cite{PrangeGirvin}
\begin{align}
n^i_{\mb{q}} = e^{-q^2 \ell_B^2/2} \sum_{a, b, k} e^{-i k q_x \ell_B^2}\hat{c}^\dagger_{k + q_y/2,  a} O^i_{ab} \hat{c}_{k - q_y/2, b}. \label{eq:rhoq}
\end{align}
On the torus the same form carries through after identifying $k \sim k +  L / \ell_B^2$, up to exponentially small terms in $\ell_B / L$.

	In the auxiliary field method the interactions are decoupled using bosonic Hubbard-Stratonovich fields ``$\phi$.'' There are a variety of possible channels for this decomposition, including the Cooper channel, but as a proof-of-principle we present here the  obvious choice $n_{\mb{q} i}$ - $n_{-\mb{q} i}$. We introduce Hermitian Hubbard-Stratonovich fields $\phi_{\mb{q} i } = \bar{\phi}_{-\mb{q} i }$ for each operator type, so that a small time step can be decomposed as
\begin{align}
e^{- d \tau  H} \sim  \prod_{i, \mb{q}'} \int d \phi_{\mb{q} i} \, \, e^{ - d \tau |\phi_{\mb{q} i} |^2 +   d \tau \sqrt{-U^i(\mb{q}) / V} \left( n^i_{\mb{q}} \bar{\phi}_{\mb{q}i} + h.c. \right) } 
\end{align}
up to normalization and the usual Trotter errors. Note that because of LL-projection, $[n^i(\mb{r}), n^i(\mb{r}')] \neq 0$. Multiplying over imaginary time-steps and integrating out the fermions, we obtain an auxiliary field path integral of the general form
\begin{align}
Z &= \mbox{Tr} \left( e^{-\beta H} \right) = \int \mathcal{D}[\phi] e^{-S[\phi]} M[\phi]\\
S[\phi] &=  \sum_{i, \mathbf{q}}  \int d \tau   \,\, |\phi_{\mb{q}i}(\tau)|^2 
\end{align}
where $M[\phi]$ is the fermion determinant for auxiliary field space-time  configuration $\phi_{\mb{q}}(\tau)$. 

The problem is sign free if $M[\phi] \geq 0$ for all $\phi$. A sufficient criteria for a sign-free determinant is the existence of two anti-unitary symmetries $T_1, T_2$ such that $T_1^2 = T_2^2 = -1$ and $T_1 T_2 = - T_2 T_1$. The symmetry must exist for \emph{any} auxiliary field configuration.\cite{White89, Assaad2008, LiYao2018}
Time-reversal is broken by the magnetic field, but at half-filling there is an anti-unitary particle-hole operation $\ph$ which exchanges empty and filled states of the LL:
\begin{align}
\ph \,\, \alpha \psi_a(r) \,\, \ph^{-1} &= \bar{\alpha} \psi_a^\dagger(r) \\
\ph \hat{c}_{k a } \ph^{-1} &= \hat{c}^\dagger_{k a}
\end{align}
For Hermitian $n^i$, $\ph$ acts  as $\ph \, n^i(\mb{r}) \, \ph^{-1} = - n^i(\mb{r})$, or $\ph \, n^i_{\mb{q}} \, \ph^{-1} = - n^i_{-\mb{q}}$.
(If $\mbox{Tr}(O^i) \neq 0$, the operators first need to be shifted according to $n^i(\mb{r}) \to n^i(\mb{r}) - \mbox{Tr}(O^i)$; we leave this shift implicit).
$\ph$ can be combined with a unitary transformation $X$ acting on the flavor index, and we will take $T_g = X_g \ph$, $g = 1,2$. The symmetry condition is
\begin{align}
O^i &=  \text{sign}\left( U^i(\mb{q}) \right)    X_g O^i X_g^\dagger
\end{align}

For a repulsive channel ($U^i > 0$), $O^i$ must be even under $X_g$, while for an attractive one ($U^i < 0$), $O^i$ must be odd.
To be sign-free, we must have $T_g^2 = X_g X_g = - 1$ and $T_1 T_2 = X_1 X_2 = - X_2 X_1 = -T_2 T_1$.
Note that $\ph$ is different than time-reversal in this respect; the first condition can always be satisfied by a phase redefinition $X_g \to i X_g$.

For the problem at hand it seems we have $O^i = \Gamma^i$, $i = 1, \cdots, 5$ with $U^i(\mb{q}) = -u_i < 0$, plus the density channel $O^0 = \mathds{1}$. But with this decomposition it is  impossible to find the two required $X_g$, because the $\Gamma$ are by definition a maximally anti-commuting set. Fortunately, for contact interactions we may use a Fierz identity  (see appendix~\ref{app:fierz}) and instead consider \cite{Wu2014}
\begin{align}
\mathcal{H} &= \frac{g}{2} (\psi^\dagger \psi)^2   + \frac{1}{2} \sum_{\mu = x, y, z}   g_\mu  (\psi^\dagger \tau^\mu \psi)^2 
\label{eq:Hg}
\end{align}
where $g = U + u_N$, $g_x = -u_N - u_4$, $g_y = - u_N - u_5$, and $g_z = 2u_N$. 
The region of interest is $g, g_z > 0$ and  $g_x, g_y < 0$. 
Decomposing in the density channels associated to these $g$, it is now easy to verify that $X_1 = i \tau^z \sigma^x$ and $X_2 = i \tau^z  \sigma^y$ satisfies the sign free condition. To handle the $\SO4$ case, we can reduce $|g_4|$ from it's $\SO5$ value.

The sign-free condition can  be seen more explicitly from Eq.~\eqref{eq:Hg} because the determinant $M$ factorizes by spin, $M[\phi] = M_\uparrow[\phi] M_\downarrow[\phi]$. This is because the $O^i$ are all diagonal in spin along direction $\sigma^z$, so the densities decompose as $n^i = n^i_\uparrow + n^i_\downarrow$, $[n^i_\uparrow, n^j_\downarrow] = 0$.
The spin-exchanging anti-unitaries $T_g$ ensure $M_\uparrow = M^\ast_\downarrow$, so the partition function can be evaluated by restricting to the $\uparrow$ orbitals,
 \begin{align}
 Z & = \int \mathcal{D}[\phi] e^{-S[\phi]} | M_{\uparrow}[\phi]|^2
 \end{align}
This restriction reduces the dimension of the linear algebra routines from $4 N_\phi \to 2 N_\phi$. 
 
Note the same reasoning carries through for projector (zero-temperature) DQMC. Restricting to spin $\uparrow$, an admissible starting state $\ket{\Omega}$ is a single filled-LL pointing along an arbitrary direction in valley-space.

\subsection{Implementation}
	The  structure of the determinant is rather different than the Hubbard model's, so we  discuss and demonstrate a naive implementation of the DQMC as a proof of principle. An optimized large-scale implementation will be presented in future work.
		
We first analyze the number of fields $\phi^i_{\mb{q}}$ required for each time-step.
On an $L \times L$ torus pierced by $N_\phi = L^2 / 2 \pi \ell_B^2$ flux, the fields $n^i_{\mb{q}}$ in-principle run over the infinite set of momenta $\mb{q} \in \frac{2 \pi}{L} (m, n)$ (though only $N_\phi^2$ of these are linearly independent). 
However, from Eq. \eqref{eq:rhoq} we see that the interaction strength is effectively
\begin{align}
U^i_{\textrm{LLL}}(\mb{q}) &\equiv U^i(\mb{q}) e^{-\frac{1}{2} q^2 \ell_B^2}
\end{align}
Thus the component of the interaction with $q < \Lambda \ell_B^{-1}$ is cut off and we can safely keep only $N_q = N_\phi \Lambda$ of the modes with an error that decreases exponentially with $\Lambda$.
There are now $\mathcal{O}(N_\phi)$ auxiliary fields per time-slice, just as there would be in the Hubbard model. 

	However, in contrast to the real-space density operators of the Hubbard model, the single-particle operators $n^i_{\mb{q}}$ are full-rank.
For low-rank operators the Sherman-Morrison formula can be used to update each of the $N_\phi$ HS fields in time $N_\phi^2$, while for a generic full-rank update takes time $N_\phi^3$. 
As a result, for inverse temperature $\beta$ and  system size $V = 2 N_\phi$, the highly naive implementation we demonstrate here (a brute force recalculation of the determinant via an LU decomposition!) scales as $\mathcal{O}(\beta V^4)$ per sweep, while the usual Hubbard DQMC scales as $\mathcal{O}(\beta V^3)$.\cite{White89}     Although slow, this  single spin flip   DQMC  allows for  the use of discrete fields  and can be implemented in the ALF package. \cite{ALF_v1}  
	
However,  the $n^i_{\mb{q}}$ do have special structure - they can be diagonalized by a fast-Fourier transform -  which allows for matrix-vector products in time $N_\phi \log(N_\phi)$ rather than the generic $N_\phi^2$.   This has for consequence that forces required for a Langevin or hybrid Monte Carlo step  \cite{Duane85}  can be computed at a cost set by   $\mathcal{O}(\beta V^3)$.   At face value  Langevin and hybrid Monte Carlo sampling  seem more efficient, but can suffer from singularities in the forces as well as ergodicity issues. \cite{Beyl17} A detailed analysis of the most efficient way to  implement the DQMC for the present problem is left for future work.

\begin{figure}
\centering
\includegraphics[width=0.9\columnwidth]{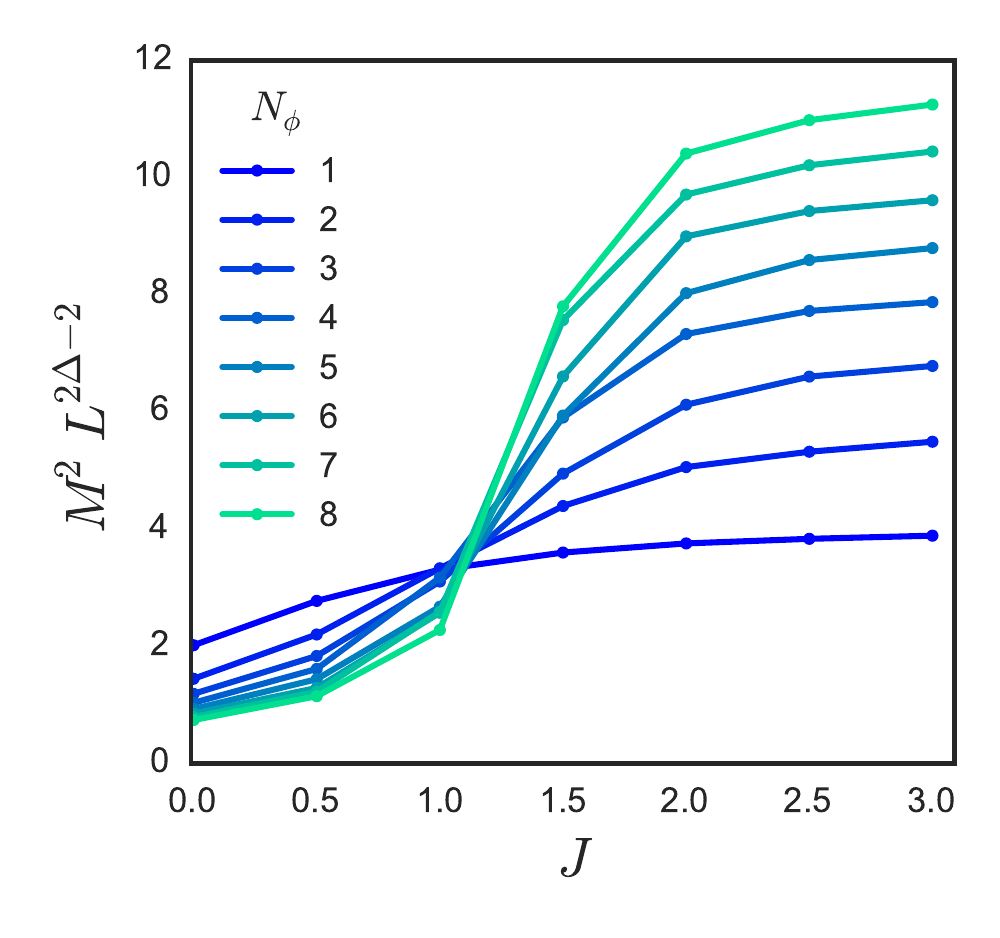}
\caption{ DQMC result for the embedding of the transverse field Ising model into the half-filled Landau level.
We plot the squared Ising magnetization $M^2$ at transverse field $h = 0.1$, as a function of the Ising coupling $J$. Data is scaled by $L^{2 \Delta - 2}$, where $\Delta \sim 0.259$ is the known scaling dimension of the Ising magnetization. As expected, the data shows a crossing around $J \sim 1$.
\label{fig:isingMC}}
\end{figure}

	As a simple test of the proposal we consider the 2+1D transverse field Ising model, which can be embedded into the $N=4$ model by choosing $U = 0, u_1 = u_2 = u_3 = u_4 = 0$, $u_5 = J$, and  introducing an additional transverse field $h \psi^\dagger \tau^x \psi$  (fields along $\tau^{x/y}$ preserve the sign-free condition).
The  ratio $h / J$ should tune an  ordered-disordered transition with Ising order parameter $M = \langle \mathbf{n}^5 \rangle$.

We present the results of a small projector DQMC simulation in Fig.~\ref{fig:isingMC}. Rather than extrapolate to zero-temperature, we evolve a transverse-polarized state $\ket{\rightarrow}$ to finite $\beta \propto \sqrt{N_{\phi}}$, e.g. $\ket{\beta} = e^{-\beta \hat{H}} \ket{\rightarrow}$.
In units with $\ell_B = \frac{e^2}{4 \pi \epsilon_0 \ell_b} = 1$, we take $\beta = 5 \sqrt{N_\phi}$ and $\Delta \tau = 0.25$ with a 2nd-order Trotter decomposition and measure the total magnetization-squared $M^2 = \langle \left (   \mathbf{n}^5_{\mathbf{q}=0} \right) ^2 \rangle_{N_\phi}$.
After scaling the data shows the crossing predicted by an Ising transition.

In addition, we have also checked the energy against exact diagonalization for $N_\phi \leq 3$ for both the Ising model and for $\SO5$-symmetric $u_i$.
	
\section{Discussion\label{sec:discussion}}

	In this work we have discussed how several 2+1D quantum phase transitions, including  deconfined quantum critical points, can be realized in half-filled continuum Landau levels which  exactly preserve  internal and spatial symmetries which would otherwise  be realized only in the IR.
The approach can be understood as a fully continuum regularization of an $\O{N}$ non-linear sigma model. 
These models can be studied using DMRG, and despite the broken time-reversal, sign-free determinantal quantum Monte Carlo.

	To our knowledge, our DMRG results are the first study of a putative DCQP beyond QMC.
While the DMRG system size ($L = 12 \ell_B$) is  much smaller than previous QMC results, we do see behavior in rough agreement. Specifically, in QMC exponents  drift slowly with system size ($\Delta_V$ flows downward) indicating  that the transition is either weakly first-order or has unconventional corrections to scaling.\cite{NahumPRX2015}
While we cannot detect such a finite-size drift given our small $L$,  we do observe a complementary phenomenon.
Our model allows us to tune a parameter, the stiffness $U$, which (if the DCQP exists) should be irrelevant.
The estimate of the scaling dimension $\Delta_V$ instead changes with $U$; for large $U$,   $\Delta_V$ is reduced and  eventually violates the unitarity bound before the transition becomes clearly first-order.
The largest value we observe is $\Delta_{V} \sim 0.7$, or $\eta \sim 0.4$.
This estimate still slightly violates the best bounds from conformal bootstrap when assuming $\SO5$ symmetry.\cite{Poland2018, Nakayama2016, Pufu2018}

Going forward, the crucial question is whether  sign-free DQMC simulations will be able to reach the system sizes required to shed new light on this issue. If so, the continuum realization may have significant advantages because we can directly identify the NLSM stiffness, $\SO5$ vector operator, and symmetric-tensor perturbations without tuning. This should greatly simplify the scaling analysis in order to investigate, for instance, whether a nearby non-unitary CFT generates a conformal window at scales above the first-order transition.\cite{Nahum2015, Wang2017, gorbenko2018walking, serna2018emergence} The stiffness $U$ could be used to control how long the flow stays in the conformal window.

A second question is which other CFTs might be realized in this fashion. When half-filling $N=4$ LLs, we have a sign-free realization of the $O(M)$ Wilson-Fisher fixed point for $M = 1, 2, 3$ and $O(M)$ DCQPs for $M = 4, 5$. It will be interesting to investigate what other models are sign-free when using $N > 4$LLs, or even attacking 5-dimensional CFTs using the quantum Hall effect in $4+1$ dimensions.

\acknowledgements
We are indebted to conversations with Snir Gazit, Tarun Grover, Luca  Iliesiu, Max Metlitski, Silviu Pufu,  Senthil Todadri,  and Ashvin Vishwanath.
MI was supported by DOE BES grant DE-SC0002140. MPZ was funded  by DOE BES Contract No.\ DE-AC02-05-CH11231,  through the Scientific Discovery through Advanced Computing (SciDAC) program (KC23DAC Topological and Correlated Matter via Tensor Networks and Quantum Monte Carlo).
FFA thanks the German Research Foundation (DFG) through the grant No. AS120/15-1.

\appendix
\begin{widetext}

\section{Equivalent parametrizations of \texorpdfstring{$\SU4$}{SU4} anisotropies\label{app:fierz}}

Here we review  the Fierz identities used to relate the two parametrizations of $\SU4$ anisotropies used in this paper, e.g., that \eq{Dirac} is equivalent to 
$$
\mathcal H = \mathcal H_0 + \frac{1}{2} \sum_{\mu=x,y,z} g_\mu (\psi^\dagger \tau^\mu \psi)^2\;,
$$
with $g_x = g_y \equiv g_\perp$. This parametrization allows a more direct conversion to experimental parameters \cite{young2014, zibrov2018}, and is crucial in the implementation of sign-free determinant quantum Monte Carlo in Sec.~\ref{sec:dqmc}.  

The equivalence can be proven by making use of a version of the Fierz identities, which we derive in the following.
We start by considering the set of matrices 
$\{O^i\} = \{\sigma^a \tau^b\}$, with $a,b\in \{0,1,2,3\}$.
These form a basis of $4\times4$ matrices.
Therefore the tensor products $\{O^i\otimes O^j\}$ form a basis of $16\times 16$ matrices, and one can perform the following decomposition:
\begin{equation}
O^i_{\alpha \beta} O^i_{\gamma \delta} = \sum_{j,k} b_{ijk} O^j_{\alpha\delta} O^k_{\gamma\beta} \;,
\label{eq:pre_fierz}
\end{equation}
where the greek indices run over electron flavors and $b$ is a matrix of coefficients.
We insert $O^m_{\delta \alpha} O^n_{\beta\gamma}$ on both sides of \eq{pre_fierz} and contract all flavor indices, obtaining
$$
\text{Tr} (O^i O^n O^i O^m) = \sum_{j,k} b_{ijk} \text{Tr}(O^jO^m) \text{Tr} (O^k O^n) \;.
$$
The $O^i$ are trace-orthogonal, with $\text{Tr}(O^i O^j) = 4\delta_{ij}$.
Moreover, any two $O$ operators either commute or anti-commute, and each $O$ squares to the identity.
Using these facts we obtain
$$
b_{imn} = \frac{1}{16} \text{Tr} (O^i O^m O^i O^n) 
= \pm \frac{1}{16} \text{Tr} (O^m O^n) 
= \pm \frac{1}{4} \delta_{mn}\;,
$$
with the $\pm$ sign decided by whether $O^i$ and $O^m$ commute or anti-commute.
We can finally rewrite \eq{pre_fierz} as
\begin{align}
(\psi^\dagger(x) O^i \psi(x))^2 
& = - \sum_j b_{ij} (\psi^\dagger(x) O^j \psi(x))^2 \;,
\label{eq:fierz} \\
b_{ij} 
& = \left\{
\begin{aligned}
& +1/4  \text{ if } O^iO^j = O^j O^i \;, \\
& -1/4 \text{ if } O^j O^i = -O^i O^j \;.
\end{aligned}
\right. \label{eq:bmatrix}
\end{align}
The extra sign comes from the Fermi statistics of the $\psi$ operators.

Direct application of \eq{fierz} shows that 
$$
(\psi^\dagger \tau^z \psi)^2 - \sum_{a=4,5} (\psi^\dagger \Gamma^a \psi)^2  + (\psi^\dagger \psi)^2 
= - (\psi^\dagger \tau^z \psi)^2  - \sum_{a=1,2,3} (\psi^\dagger \Gamma^a \psi)^2 \;,
$$
which implies
\begin{equation}
(\psi^\dagger \tau^z \psi)^2 
= -\frac{1}{2}(\psi^\dagger \psi)^2 - \frac{1}{2} \sum_{a=1,2,3} (\psi^\dagger \Gamma^a \psi)^2  + \frac{1}{2} \sum_{a=4,5} (\psi^\dagger \Gamma^a \psi)^2  
\end{equation}
This identity allows us to map the two parametrizations:
\begin{equation}
\frac{V}{2} (\psi^\dagger \psi)^2 + \frac{g_\perp}{2} \sum_{\mu = x,y} (\psi^\dagger \tau^\mu \psi)^2 + \frac{g_z}{2} (\psi^\dagger \tau^z \psi)^2 =
\frac{U}{2} (\psi^\dagger \psi)  - \frac{u_N}{2} \sum_{a=1,2,3} (\psi^\dagger \Gamma^a \psi)^2 - \frac{u_K}{2} \sum_{a=4,5} (\psi^\dagger \Gamma^a \psi)^2 
\label{eq:parametrizations}
\end{equation}
with 
\begin{equation}
U = V - \frac{1}{2} g_z,\quad
u_N = \frac{1}{2} g_z,\quad
u_K = -g_\perp - \frac{1}{2} g_z\;.
\end{equation}

\end{widetext}

\bibliography{graphene_deconfined}

\end{document}